\documentclass[pra,aps,twocolumn,superscriptaddress,10]{revtex4-2}
\usepackage[english]{babel}
\usepackage{amsmath,amssymb,amsfonts,bbold}

\usepackage{graphicx}
\usepackage[colorlinks=True,linkcolor=blue,citecolor=blue,urlcolor=blue]{hyperref}
\usepackage{import}
\usepackage{booktabs}

\usepackage[dvipsnames]{xcolor}

\makeatletter
\renewcommand*{\fnum@figure}{{\normalfont\bfseries \figurename~\thefigure}}
\renewcommand*{\fnum@table}{{\normalfont\bfseries \tablename~\thetable}}
\makeatother

\newcommand{\dd}{\text{d}}
\newcommand{\ii}{\text{i}}
\newcommand{\ee}{\text{e}}

\begin{document}
\title{An Algebraic Approach to Bifurcations in Kerr Ring and Fabry-Pérot Resonators}
\author{Juan Diego Mazo-Vásquez}
\email{juan-diego.mazo-vasquez@mpl.mpg.de}
\affiliation{Max Planck Institute for the Science of Light, Staudtstraße 2, 91058 Erlangen, Germany}
\affiliation{Department of Physics, Friedrich-Alexander Universität Erlangen-Nürnberg, Staudtstraße 7, 91058 Erlangen, Germany}
\author{Julius T. Gohsrich}
\email{julius.gohsrich@mpl.mpg.de}
\affiliation{Max Planck Institute for the Science of Light, Staudtstraße 2, 91058 Erlangen, Germany}
\affiliation{Department of Physics, Friedrich-Alexander Universität Erlangen-Nürnberg, Staudtstraße 7, 91058 Erlangen, Germany}
\author{Flore K. Kunst}
\email{flore.kunst@mpl.mpg.de}
\affiliation{Max Planck Institute for the Science of Light, Staudtstraße 2, 91058 Erlangen, Germany}
\affiliation{Department of Physics, Friedrich-Alexander Universität Erlangen-Nürnberg, Staudtstraße 7, 91058 Erlangen, Germany}
\author{Lewis Hill}
\email{lewis.hill@mpl.mpg.de}
\affiliation{Max Planck Institute for the Science of Light, Staudtstraße 2, 91058 Erlangen, Germany}

\begin{abstract}
%High-quality Kerr resonators are a key platform for studying nonlinear optical phenomena, where bifurcations such as optical bistability and spontaneous symmetry breaking are both oftheoretical and practical significance.
Nonlinear phenomena such as optical bistability and spontaneous symmetry breaking play a central role in Kerr resonators, and are increasingly exploited in photonic integrated circuits for all-optical information processing. In this work, we present an analytical framework allowing to find the stationary states and their bifurcations for the propagating fields in Kerr ring and Fabry-Pérot resonators, which can be generalized to other nonlinear systems. Using tools from nonlinear algebra, namely, polynomial resultants and Gröbner bases, we derive compact polynomial expressions describing the system's full solution in both intensity and amplitude representations. The bifurcations follow directly from these expressions, and are additionally characterized as exceptional points of an auxiliary linear non-Hermitian system. Together, these results unify optical bistability and spontaneous symmetry breaking within a single analytical framework, and offer a route toward improved control of nonlinear optical systems, and the design of photonic devices.
%This work unifies key phenomena in Kerr resonators under the broader framework of nonlinear algebra and offers better control of nonlinear optical systems and the design of photonic devices -- enabled by full analytic control.

\end{abstract} 

\maketitle

\section{Introduction}

\begin{figure}[hbt!]
    \centering
    \includegraphics[width=\linewidth]{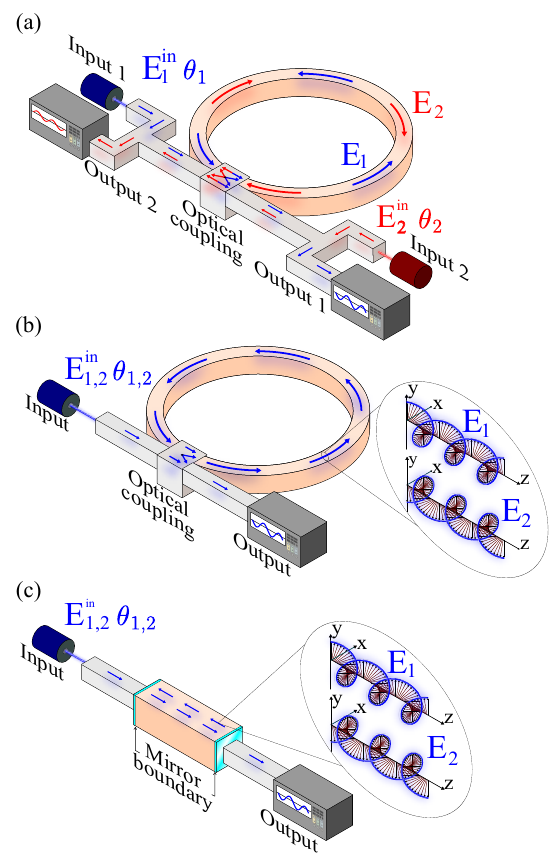}
    \caption{\textbf{Kerr ring and Fabry-Pérot resonator configurations.} Kerr ring configurations: (a) two laser beams counter-propagate within the resonator after entering through an optical coupler, and (b) a single laser beam pumps elliptically polarized light into the resonator, with right- and left-circularly polarized components co-propagating within it. (c) Fabry-Pérot resonator configuration, where elliptically polarized light circulates back and forth due to reflections at the boundaries of the cavity. In all three scenarios, the output intensity is measured after leaving the resonator.}
    \label{fig:fig1}
\end{figure}

High-quality (high-Q) optical resonators have emerged as a powerful platform for exploring nonlinear optical phenomena. Because nonlinear light-matter interactions are generally much weaker than linear ones, long interaction lengths are typically required to observe or exploit nonlinear effects. High-Q resonators mitigate this limitation by confining photons to circulate through the same waveguide structure many times, effectively emulating a long nonlinear medium within a compact footprint. Although Fig.~\ref{fig:fig1} illustrates a more complex multi-field resonator configuration, it includes a ring resonator and a Fabry-Pérot cavity exemplifying this principle. In this work, we focus on high-Q Kerr resonators, which are fabricated from materials that, in addition to a linear optical response, exhibit a dominant third-order nonlinear polarization response characterized by the susceptibility $\chi^\mathrm{(3)}$. Common materials are silicon nitride~\cite{YeOL2019}, silica~\cite{Murnieks2023}, and lithium niobate~\cite{WangNatCommun2019}. 

One of the attractions of Kerr resonators stems from the fact that theoretical and experimental progress goes hand-in-hand. They offer a versatile and accurate platform for realizing complex dynamical systems, facilitating fundamental investigations into nonlinear optical phenomena. Moreover, their practical relevance spans a broad range of engineering domains, with applications including all-optical computing, precision metrology, and integrated photonic circuits. Many of the nonlinear phenomena observed in Kerr resonators, such as breathing temporal cavity solitons~\cite{XuOL2022} and localized structures~\cite{SanvertCLEO2025}, the recently reported faticons~\cite{LucasPRL2025,HillCLEO2025}, advanced frequency combs based on soliton crystals~\cite{CampbellOE2024}, integrated optical isolators and circulators~\cite{DelBinoOptica2018}, logic gates for all-optical computing~\cite{GhoshLPR2025}, optical switches~\cite{ZhangPR2025}, and random number generators~\cite{QuinnOL2023}, are either enhanced by, or fundamentally dependent on, two core dynamical effects: optical bistability and spontaneous symmetry breaking (SSB). 

Optical bistability, a well-known phenomenon in nonlinear systems~\cite{AgrawalPRA1979, KaplanOL1981, GoldstonePRL1984, RukhlenkoOL2010, KaplanOC1982, VanIEEE2002}, occurs when two steady-state solutions exist for the same experimental parameters.
SSB is a phenomenon in which two initially identical system properties develop an asymmetry following an infinitely small change to some system parameter. In the context of Kerr resonators, it occurs above a certain threshold, due to the intensity-dependent refractive index, which amplifies small perturbations, breaking the initial symmetry and stabilizing an asymmetric state~\cite{WoodleyPRA2018}.
Both phenomena, associated with bifurcations of the stationary states of the system, result in hysteresis effects and are critical for applications such as manipulating multiplexing of light in integrated circuits~\cite{GhoshPR2024}, controlling the light polarization of continuous wave lasers~\cite{MoroneyNatCom2022}, or enhancing sensitivity of Sagnac interferometers~\cite{SilverOptica2021}.

We show pictorial representations of two Kerr ring and one of a Fabry-Pérot resonator in Figs.~\ref{fig:fig1}(a,b), and (c), respectively. In Fig.~\ref{fig:fig1}(a), two counter-propagating, elliptically polarized pump lasers are injected into the cavity. In Fig.~\ref{fig:fig1}(b), a single elliptically polarized pump laser is injected and subsequently analyzed in terms of its decomposed left- and right-circularly polarized components. Figure~\ref{fig:fig1}(c) depicts a Fabry-Pérot resonator, at which elliptically polarized light propagates back and forth due to the multiple reflections at the boundaries of the cavity.

The dynamics of the propagating fields in the scenarios shown in Fig.~\ref{fig:fig1} arise from the intricate interplay between cavity losses, the input pump (gain), dispersive, and nonlinear effects. The Lugiato-Lefever equation (LLE)~\cite{LugiatoPRL1987}, in its purely temporal form~\cite{HaeltermanOC1992}, provides an excellent framework for describing the propagation dynamics of a single field in a Kerr ring resonator, with an adaptation extending its applicability to Kerr Fabry-Pérot cavities~\cite{ColePRA2018, MoroneyNatCom2022}. Configurations such as those shown in Fig.~\ref{fig:fig1}(a-c) are modeled using coupled LLEs for counter-propagating fields~\cite{WoodleyPRL2021}, for co-propagating fields with orthogonal polarizations~\cite{XuNatCom2021} and non-degenerate frequency modes~\cite{TrinchaoArXiv2026}, and for a Kerr Fabry-Pérot cavity with orthogonal polarizations~\cite{HillComPhys2024, MaiCommPhys2024, CampbellArXiv2025}, respectively. In the scenario of two elliptically polarized counter-propagating beams, multi-stage symmetry breaking might occur due to the interaction of the four fields (when decomposing the counter-propagating beams in right- and left-circularly polarized beams) circulating within the resonator~\cite{HillComPhys2023}.

It is well-established that, under homogeneous stationary conditions, sufficient to observe both optical bistability and SSB, all three systems have their circulating field envelope amplitudes $E_{1,2}$ governed by the same algebraic relations~\cite{HillarXiv2025}:
\begin{equation}\label{eq:eqsfieldsgeneral}
    E^\mathrm{in}_{1,2} - E_{1,2} +\ii (-\theta_{1,2} + A |E_{1,2}|^2 + B|E_{2,1}|^2)E_{1,2}=0,
\end{equation}
\noindent which, upon multiplication by their complex conjugate, yield a common equation for the circulating field intensities $P_{1,2} = |E_{1,2}|^2$:
\begin{equation} \label{eq:eqsintensitiesgeneral}
    P^\mathrm{in}_{1,2} - \left[1+(-\theta_{1,2} + A P_{1,2} + B P_{2,1})^2\right]P_{1,2} = 0.
\end{equation}
Here, $\theta_{1,2}$ are the cavity detunings for each field, defined as the difference between the pump laser frequency and the nearest cavity resonance; $E^\mathrm{in}_{1,2}$ are the input-pump-field amplitudes, such that $P_{1,2}^\mathrm{in} = |E^\mathrm{in}_{1,2}|^2$ are the pumping field intensities; and $A$ and $B$ are the self- and cross-phase modulation coefficients, respectively, determined by the components of the third-order susceptibility tensor~$\chi^\mathrm{(3)}$~\cite{HillPRA2020}, and we assume $A$ and $B$ to be constant, real and positive.

While the connection between optical bistability, SSB and bifurcation theory is already established~\cite{WoodleyPRA2018, BithaPRE2023}, the complexity of Eqs.~\eqref{eq:eqsfieldsgeneral} and \eqref{eq:eqsintensitiesgeneral} has so far prevented full analytical access to these systems.
The main purpose of this work is to fill this gap by applying techniques from nonlinear algebra~\cite{Gelfand1994, Cox2015}, namely, polynomial resultants and Gröbner bases, to
derive analytical expressions encoding the intensities and amplitudes of the homogeneous stationary states of the LLE. These expressions also give access to the bifurcations describing optical bistability and spontaneous symmetry breaking, and we can associate these bifurcations with exceptional points (EPs) -- defective points at which the eigenvalues and eigenvectors coalesce -- of an auxiliary linear non-Hermitian system.

\section{Results and discussion}\label{section:Results}

Even though Eq.~\eqref{eq:eqsfieldsgeneral} depends on complex amplitudes, for both propagating field amplitudes and pumping fields, Eq.~\eqref{eq:eqsintensitiesgeneral} depends solely on real variables. The physical solutions to Eq.~\eqref{eq:eqsintensitiesgeneral} are the real and positive values of $P_{1,2}$, corresponding to the measurable circulating intensities within the resonator. Two distinct regimes emerge depending on the control parameters $\theta_{1,2}$ and $P^\mathrm{in}_{1,2}$: Balanced input parameters, where $\theta_1 = \theta_2$ and $P^\mathrm{in}_1 = P^\mathrm{in}_2$, and imbalanced input parameters, where either $\theta_1 \neq \theta_2$ or $P^\mathrm{in}_1 \neq P^\mathrm{in}_2$, or both unequal. The following sections analyze these two regimes separately.

\subsection{Balanced Input Parameters: Intensities}\label{section:equal}

\begin{figure*}
    \centering
    \includegraphics[width=\linewidth]{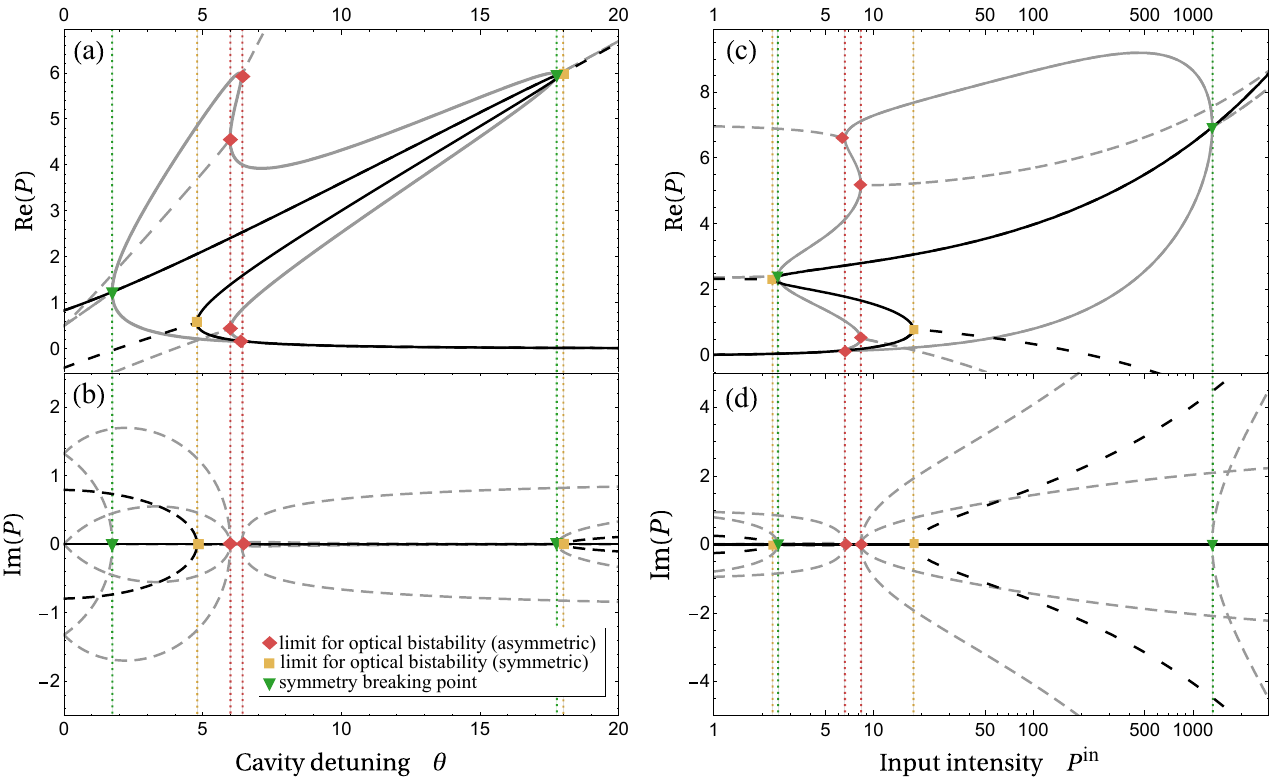}
    \caption{\textbf{Real and imaginary parts of the circulating intensities.} Real (a,c) and imaginary (b,d) parts of the roots $P$ of $p(P)$ as functions of the cavity detuning in (a,b) with $P^\mathrm{in} = 6$, and the input intensity in (c,d) with $\theta = 7$. In both cases $A=1$ and $B=2$. Solid lines in (a,c) represent real, non-negative $P$ corresponding to physical solutions, while dashed lines indicate roots with nonzero imaginary part, corresponding to unphysical solutions. Symmetric and asymmetric solutions are shown in black and gray, respectively. Green triangles mark the bifurcations at the SSB points, while red diamonds and yellow squares denote bifurcations at the optical bistability limits for asymmetric and symmetric solutions, respectively. The bifurcations occur at the transition points between physical and unphysical solutions, marked by the dotted vertical lines.}
    \label{fig:fig2}
\end{figure*}

We start with the balanced configuration due to its simplicity and high relevance, and focus on the intensities $P_{1,2}$. Balanced control parameters are realized in Fig.~\ref{fig:fig1}(a) by injecting two identical pump lasers into the resonator, differing only in their propagation direction, while in configurations (b) and (c), a single linearly polarized pump laser is injected. Under the conditions $P^\mathrm{in}_1 = P^\mathrm{in}_2=P^\mathrm{in}$ and $\theta_1=\theta_2=\theta$, Eq.~\eqref{eq:eqsintensitiesgeneral} reads
\begin{equation} \label{eq:intensitiesbalanced}
P^\mathrm{in} - \left[1 + \left(-\theta + A P_{1,2} + B P_{2,1} \right)^2 \right] P_{1,2} = 0,
\end{equation}
which has been widely studied for its ability to exhibit both SSB and optical bistability. At the SSB point, a symmetric solution, i.e., $P_1 = P_2$, becomes unstable, giving rise to an asymmetric state, at which $P_1 \neq P_2$, under an infinitesimal perturbation. In the optically bistable region, multiple solutions for $P_{1,2}$ coexist within specific regions of parameter space~\cite{KaplanOC1982, WrightPRA1985, HillPRA2020,WoodleyPRL2021}.
Equation~\eqref{eq:intensitiesbalanced} can be written as an implicit equation, $f_{1,2}(P_1, P_2) = 0$, where $f_{1,2}$ are bivariate polynomials in the variables $P_1$ and $P_2$ satisfying the symmetry $f_1(P_1, P_2) = f_2(P_2, P_1)$, i.e., the exchange $P_1\rightleftharpoons P_2$ leaves the set of polynomials $f_{1,2}$ invariant. To eliminate one of the variables, we consider the resultant $\mathrm{res}\left(f_1,f_2, P_{2,1}\right) = p(P_{1,2})$ (cf. Methods). The resulting ninth-order polynomial $p(P_{1,2})$ contains all the possible stationary state solutions for $P_{1,2}$, such that after solving for a given variable, e.g., $P_1$, the solutions for the other variable, e.g., $P_2$, can be computed from Eq.~\eqref{eq:intensitiesbalanced}. The intensities fall into two categories: they can be either equal, at which $P_1=P_2$, which we refer to as symmetric, or unequal, at which $P_1\neq P_2$, which we refer to as asymmetric or symmetry-broken. In the former case, Eq.~\eqref{eq:intensitiesbalanced} becomes
\begin{equation}\label{eq:symmetricintensities}
    P^\mathrm{in} - \left[1+\left(-\theta+ (A+B)P)\right)^2\right] P =0,
\end{equation}
which is a third-order polynomial in $P$, where $P$ is either $P_1$ or $P_2$, and we identify this as $p^{\text{s}}(P)=0$. This corresponds to the case at which a single beam circulates within the resonator with a self-phase modulation constant $A+B$~\cite{HillPRA2020}. The polynomial $p(P)$, where $P$ is either $P_1$ or $P_2$, containing all symmetric and asymmetric solutions, can be factorized as
\begin{equation}\label{eq:polsintensities}
    p(P) = p^{\text{s}}(P)\cdot p^{\text{a}}(P),
\end{equation}
where the sixth-order polynomial $p^{\text{a}}(P)$ gives information about the asymmetric solutions $P_1\neq P_2$ (coefficients of~$p^{\mathrm{a}}$ are given in the Supplementary Information~(SI)~\cite{SI}).

The polynomials in Eq.~\eqref{eq:polsintensities} have nine roots in general, but out of these, there can be several real roots for a single set of parameters, corresponding to physical solutions.
The bifurcation points, at which a transition between physical and unphysical solutions occurs, correspond to the onset of optical bistability and SSB.
In Fig.~\ref{fig:fig2} we show the real and imaginary parts of the roots of $p^{\text{s}}$ (black lines) and $p^{\text{a}}$ (gray lines) as functions of both cavity detuning (Fig.~\ref{fig:fig2}(a,b)) and input intensity (Fig.~\ref{fig:fig2}(b,c)). Solid lines denote real-valued solutions for $P$, while dashed lines indicate solutions with non-zero imaginary components. From the SSB points, where the symmetry-broken region starts or ends, the circulating intensities split, and one follows the upper part of the so-called \textit{asymmetric bubble}, whereas the other one follows the lower part. They merge again when a second SSB point occurs, which is called the spontaneous symmetry restoration point.

The points in parameter space where bifurcations occur can be identified by analyzing at which points either the discriminant of $p^{\text{s}}$ or $p^{\text{a}}$ vanishes, i.e., when repeated roots occur. For instance, the discriminant of $p^{\mathrm{s}}$ in Eq.~\eqref{eq:symmetricintensities} with respect to $P$ is

\begin{align}\label{eq:disps}
        &\mathrm{dis}(p^{\mathrm{s}}, P) = -(A+B)^2 \left[27 (P^\mathrm{in})^2 (A+B)^2\right. \\ &\left.
        -4 (A+B) \theta ^3 P^\mathrm{in}
         -36 \theta  P^\mathrm{in} (A+B)+4 \theta ^4+8 \theta ^2+4\right], \notag
\end{align}
which vanishes whenever

\begin{equation}\label{eq:pin-optical-bistable}
    P^\mathrm{in} = \frac{2 \theta  \left(\theta ^2+9\right) \pm 2\sqrt{\left(\theta ^2-3\right)^3}}{27 (A+B)},
\end{equation}
corresponding to limits of optical bistability.
The other discriminants can be found in the Methods.

\begin{figure}
    \centering
    \includegraphics[width=\linewidth]{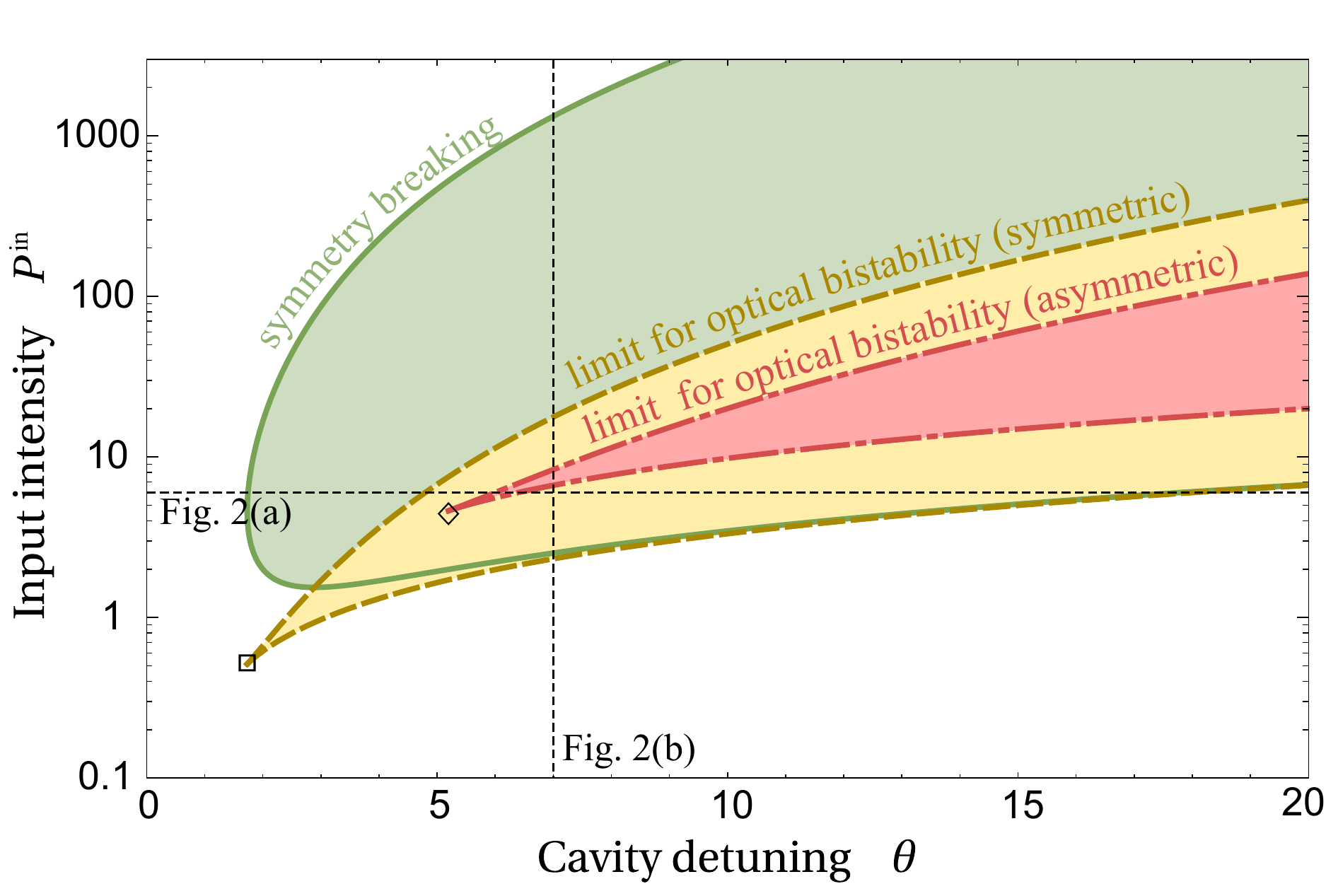}
    \caption{\textbf{Bifurcation lines in parameter space.} Bifurcation lines for SSB and optical bistability in the parameter space spanned by the cavity detuning $\theta$, and the input intensity $P^\mathrm{in}$, with $A=1$ and $B=2$. The dot-dashed red and dashed yellow lines correspond to the limits for optical bistability in the asymmetric and symmetric cases, respectively; the points at which these lines merge are marked with an empty diamond and an empty square, respectively. The solid green line corresponds to the SSB line. The black dashed lines correspond to the parameter scans in Fig.~\eqref{fig:fig2}. Green, yellow, and red shaded areas correspond to the symmetry-broken, optical bistability of the symmetric, and optical bistability of the asymmetric solution, respectively.}
    \label{fig:fig3}
\end{figure}

The different types of bifurcations are shown in Fig.~\ref{fig:fig3} in the parameter space spanned by $\theta$ and $P^\mathrm{in}$ and form lines.
The dashed yellow line corresponds to optical bistability in the symmetric case, see Eq.~\eqref{eq:pin-optical-bistable}.
The solid green and dot-dashed red lines are where the discriminant of the polynomial encoding the asymmetric solution vanishes, i.e., where \mbox{$\mathrm{dis}(p^{\mathrm{a}}, P) = 0$} (cf. Methods).
From Fig.~\ref{fig:fig2}, we can infer that these lines correspond to the  symmetry-breaking line and the limits of optical bistability in the asymmetric regime, respectively.
More details can be found in the Methods.
In Fig.~\ref{fig:fig2}, we see that at the limits of optical bistability, in both symmetric (black lines) and asymmetric (gray lines) regimes, two degenerate unphysical solutions bifurcate into two physical ones, indicated by an empty square and an empty diamond, respectively. The associated lines in the control parameter space merge at 
\begin{equation}\label{eq:EP3sym}
(\theta, P^\mathrm{in}) = \left( \pm\sqrt{3}, \frac{8}{3\sqrt{3}(A + B)} \right),
\end{equation}
for the symmetric case, whereas for the asymmetric case, the lines merge at
\begin{equation}\label{eq:EP3asymm}
(\theta, P^\mathrm{in}) = \left( \pm\frac{\sqrt{3}(A + B)}{|A-B|}, \frac{8(A + B)}{3\sqrt{3} |A-B|} \right).
\end{equation}

The $\pm$ sign accounts for positive or negative cavity detuning differences, and we only show the positive detuning case in Fig.~\ref{fig:fig3} -- the negative region is the mirror image of the positive one. The white region corresponds to the domain in which only symmetric and mono-stable solutions are present. The lines corresponding to the boundaries of optical bistability in the symmetric regime emerge from saddle-node bifurcations of the stable solution branches~\cite{BraeckeveldPRR2024}. Meanwhile, the merging points in Eqs.~\eqref{eq:EP3sym} and \eqref{eq:EP3asymm} mark cusp catastrophes of codimension two~\cite{KwongarXiv2025, Saunders1980}, where the equilibrium surfaces are described by the polynomials $p^{\text{s}}$ and $p^{\text{a}}$, respectively.

To characterize the bifurcations of the nonlinear system we employ tools from non-Hermitian physics.
Because the relevant information is encoded in the roots of~$p$, we now interpret this polynomial as the characteristic polynomial of an auxiliary non-Hermitian system $M$, so that $p(P)=\det(M-P \mathbb{1}_9)$, where $\mathbb{1}_N$ is the $N \times N$ identity matrix.
Such a mapping is not unique, and we will show its usefulness in the following.
We previously factorized $p$ into polynomials encoding the symmetric and asymmetric solutions, cf. Eq.~\eqref{eq:polsintensities}, which are two physically different and mutually exclusive situations.
Thus, we map $p^{\text{s}}$ and $p^{\text{a}}$ individually to the matrices~$M^\mathrm{a}$ and~$M^\mathrm{s}$ using the Frobenius companion matrix~\cite{EastmanLA2014} as conventional choice, so that $M = M^{\mathrm{a}} \oplus M^\mathrm{s}$.

With this mapping, we find that the bifurcation points in the parameter space spanned by $\theta$ and $P^\mathrm{in}$ correspond to EPs in $M$.
This can be verified by computing the Jordan decomposition of $M$ evaluated at the bifurcation points.
More specifically, the limits of optical bistability in the symmetric regime correspond to EPs of order two (EP2s) -- where the order refers to the number of coalescing eigenvectors -- stemming from $M^\mathrm{s}$, resulting in a Jordan block, denoted here as $J^\mathrm{s}_2$.
The limits of optical bistability in the asymmetric regime correspond to two EP2s at different intensities $P$.
We refer to this situation as $J^\mathrm{a}_2 \oplus \bar{J}^\mathrm{a}_2$, where the overline signifies that the Jordan blocks are evaluated at different $P$.
Finally, the SSB bifurcation points correspond to having a single EP2 stemming from $M^\mathrm{a}$, but there is an additional solution stemming from $M^\mathrm{s}$ at the same intensity $P$, and we denote it $J_2^\mathrm{a} \oplus J^\mathrm{s}_1$.
This provides a direct connection between the considered bifurcations and EPs, summarized in Table~\ref{table}.
For some isolated points, there might occur equal eigenvalues for $M^\mathrm{s}$ and $M^\mathrm{a}$, which corresponds to a two-fold degeneracy of $M$.
They correspond to accidental degeneracies~\cite{SayyadSciPost2023} and not to EP2s.

Importantly, in all cases, the intensities around the bifurcation point, cf. Fig.~\ref{fig:fig2}, show a square-root behavior.
This is also expected for the dispersion around an EP2~\cite{MiriScience2019, WiersigPRA2016}, further solidifying the analogy between bifurcations and EPs.
The merging points, cf. Eqs.~\eqref{eq:EP3sym} and~\eqref{eq:EP3asymm}, correspond to third-order EPs, and the intensities around those show cubic-root behavior.
Another observation in the context of the non-Hermitian analysis is the fact that all coefficients of $p$ are real.
As such, $M$ falls in the similarity class of pseudo-Hermitian systems~\cite{MontagJMP2024, MontagPRR2025}, and roots of $p$ -- equivalent to eigenvalues of $M$ -- are either real or occur in conjugate pairs.
Furthermore, due to this similarity, the EP2s do not appear as isolated points in parameter space, but rather trace out lines, as seen in Fig.~\ref{fig:fig3}.

\begin{table}[bt]
\begin{tabular}{cc}
\toprule
Bifurcation                                                                          & Jordan structure                                                  \\ \midrule
\begin{tabular}[c]{@{}c@{}}Limits of optical bistability\\  (symmetric regime)\end{tabular}  & $M = J^\mathrm{s}_2 \oplus D_7\hspace{0.73cm}$                    \\[1.2em]
\begin{tabular}[c]{@{}c@{}}Limits of optical bistability\\  (asymmetric regime)\end{tabular} & $M = J^\mathrm{a}_2\oplus \bar{J}^\mathrm{a}_2 \oplus D_5$  \\[1.2em]
 SSB points                      & $M =J_2^\mathrm{a} \oplus J^\mathrm{s}_1 \oplus D_6$\\
\bottomrule
\end{tabular}
\caption{\textbf{Jordan structure of the auxiliary non-Hermitian matrix at the EPs}.
Generic Jordan Block structure at the EPs of the auxiliary non-Hermitian matrix $M$, corresponding to steady-state bifurcations in the solution of Eq.~\eqref{eq:intensitiesbalanced}.
Here, $J_k^\mathrm{s}$~($J_k^\mathrm{a}$) denotes Jordan blocks of size $k=1,2$ and stems from $M^\mathrm{s}$~($M^\mathrm{a}$).
The overline emphasizes that the two Jordan blocks are evaluated at different intensities~$P$.
The diagonal matrix $D_k$ encodes the remaining solutions and is non-degenerate with respect to all $J$s.
}\label{table}
\end{table}

\subsection{Balanced Input Parameters: Amplitudes}
Next, we solve the amplitude equations for balanced input parameters and find their steady-state bifurcations.
Because Eq.~\eqref{eq:eqsfieldsgeneral} contains absolute value squares of the field amplitudes, we considered a slightly different approach to transform the equations to polynomial forms.
To do so, we treat the complex conjugates of the amplitudes, $E_{1,2}^*$, as an independent amplitude, $E_{1,2}^* \to F_{1,2}$, so that $|E_{1,2}|^2=E_{1,2}F_{1,2}$.
Now we write down Eq.~\eqref{eq:eqsfieldsgeneral} and its complex conjugate in terms of these auxiliary amplitudes, yielding the four coupled \emph{polynomial} equations for the four unknown amplitudes $E_1$, $E_2$, $F_1$ and $F_2$:
\begin{align}\label{eq:fields-auxiliaries}
	E^\mathrm{in}_{1,2} - E_{1,2} +\ii (-\theta_{1,2} + A E_{1,2} F_{1,2} + B E_{2,1} F_{2,1}) E_{1,2}&=0, \notag \\
    E^{\mathrm{in}*}_{1,2} - F_{1,2} -\ii (-\theta_{1,2} + A F_{1,2} E_{1,2} + B F_{2,1}E_{2,1})F_{1,2}&=0.
\end{align}

In this polynomial form, we can employ Gröbner bases to eliminate three out of the four amplitudes to obtain a univariate polynomial~\cite{Cox2005} in the remaining amplitude.
These polynomials are the amplitude analogs of Eq.~\eqref{eq:polsintensities}, and provide information about all the possible stationary states for that particular amplitude.
Having determined these stationary states, one can iteratively find all the other amplitudes~\cite{Cox2015}.
More details, and a simple example, on Gröbner bases can be found in the Methods.

\begin{figure*}[hbt!]
    \centering
    \includegraphics[width=\linewidth]{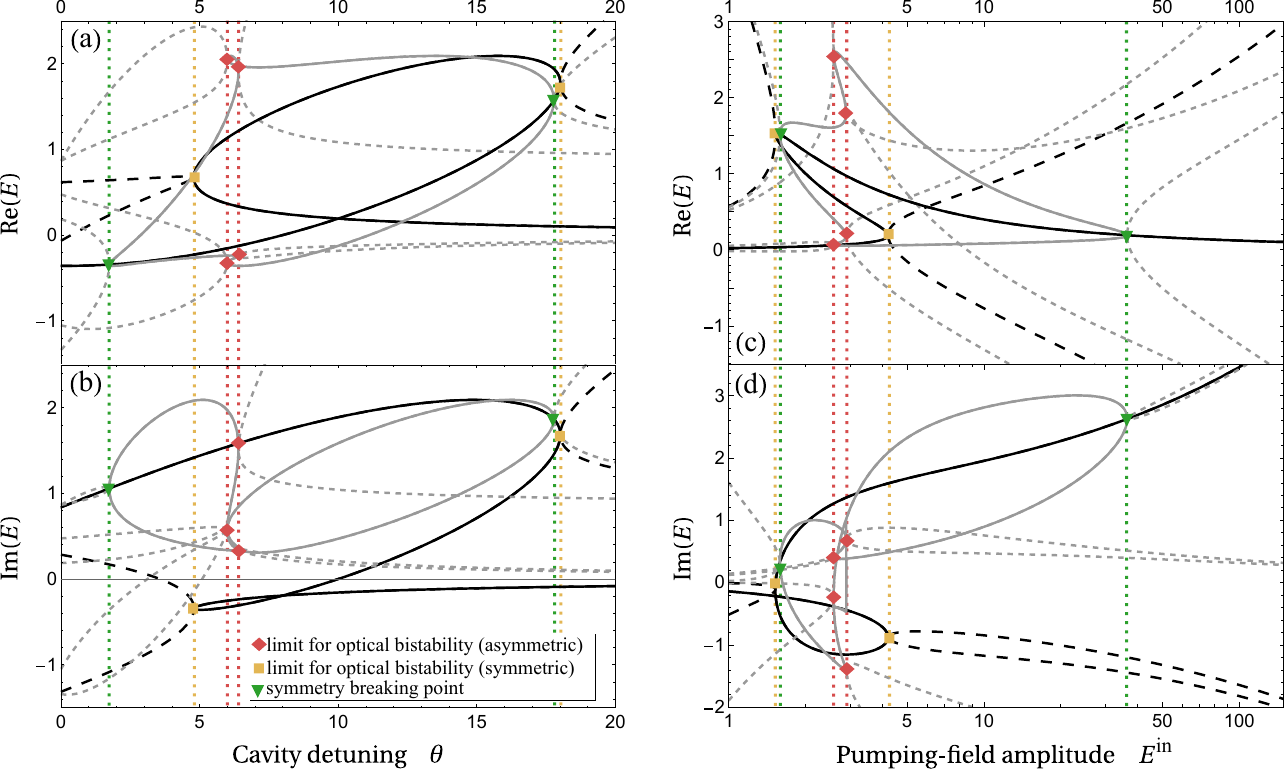}
    \caption{\textbf{Real and imaginary parts of the field amplitudes.} Same as Fig.~\ref{fig:fig2}, but for the filed amplitudes $E$, setting $E^\mathrm{in} =\sqrt{P^\mathrm{in}} \cdot \ee^{\ii \pi/4}$. The dashed lines correspond to solutions not fulfilling the physicality constraint.}
    \label{fig:fig4}
\end{figure*}

The set of polynomials in Eq.~\eqref{eq:fields-auxiliaries}, assuming $E^\mathrm{in}_1 = E^\mathrm{in}_2=E^\mathrm{in}$ and $\theta_1=\theta_2=\theta$, are symmetric under the simultaneous exchange of $E_1 \rightleftharpoons E_2$ and $F_1\rightleftharpoons F_2$. Therefore, the univariate polynomials in the variables $E_1$ and $E_2$ (or $F_1$ and $F_2$) are equal.
Again, these are ninth-order polynomials and can be written as the product of a third- and a sixth-order polynomials, encoding symmetric ($E_1 = E_2$ and $F_1 = F_2$) and asymmetric ($E_1\neq E_2$ and $F_1\neq F_2$) roots.
Eliminating $F_1$, $F_2$ and either $E_1$ or $E_2$ results in the univariate polynomial $\tilde{p}(E)=\tilde{p}^\mathrm{s}(E)\,\tilde{p}^\mathrm{a}(E)$.
As the set of polynomials in Eq.~\eqref{eq:fields-auxiliaries} is symmetric under $E_{1,2}^* \rightleftharpoons F_{1,2}$, eliminating $E_1$, $E_2$ and either $F_1$ or $F_2$ gives the conjugate polynomial, i.e., $\tilde{p}^*(F)$.
All coefficients of the polynomials can be found in the SI.

As anticipated, the physical solutions to Eq.~\eqref{eq:fields-auxiliaries} need to satisfy the conjugation symmetry $E_{1,2}^*=F_{1,2}$.
This also ensures that the associated circulating intensities, $P_{1,2}=|E_{1,2}|^2=E_{1,2}F_{1,2}$, are real and correspond to the physical solutions in the intensity picture.

The bifurcations, computed from the discriminants of the univariate polynomials, occur at the same parameter values as the bifurcations associated to the limits for optical bistability and SSB observed in the intensity-based analysis (cf. Methods).
The bifurcation analysis in terms of EPs of an auxiliary matrix $\tilde{M}$ is analogous to the discussion on the intensities.
We also want to emphasize, that even though the phase of the pumping field, $\phi=\mathrm{arg}(E^\mathrm{in})$, adds a degree of freedom to the parameter space, the bifurcation points are independent of it (cf. Methods).

Figure~\ref{fig:fig4} presents the real (a,c) and imaginary (b,d) parts of the field amplitudes. Panels (a,b) show a detuning scan, while panels (c,d) show intensity scans, analogous to Fig.~\ref{fig:fig2}. Physically valid solutions, those satisfying the conjugation symmetry, are depicted with solid lines, whereas unphysical solutions violating this symmetry appear dashed. Again, gray and black lines correspond to symmetric and symmetry-broken solutions, respectively.

\subsection{Imbalanced Input Parameters}\label{section:imbalanced}

\begin{figure*}
    \centering
    \includegraphics[width=\linewidth]{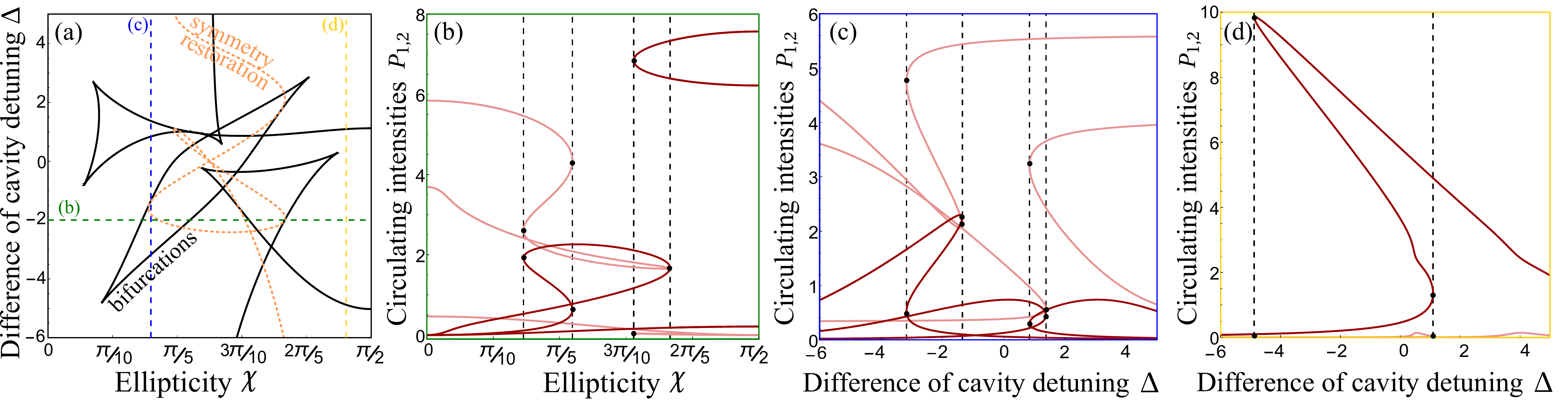}
    \caption{\textbf{Parameter space spanning imbalanced parameters, and different parameter scans.} (a) Parameter space spanned by the ellipticity angle $\chi$ and the difference of cavity detuning $\Delta$, at which bifurcations occur (solid black lines). The dotted orange line corresponds to the points at which the symmetry between the intensities is recovered, i.e., where $P_1 =P_2$. Dashed lines correspond to the parameter scans in (b-d). Circulating intensities as functions of (b) the ellipticity for $\Delta = -2$, (c,d) as functions of the difference of cavity detuning with $\chi = 0.5$, and $\chi=1.45$, respectively. Light- and dark-red lines correspond to the circulating intensities. The dashed black lines and black dots in panels (b-d) correspond to the position of the bifurcations in (a). All panels assume $A=1$, $B=2$, $\theta=5$ and $P^\mathrm{in}=10$.}
    \label{fig:fig5}
\end{figure*}

We now consider the more general situation, in which the two propagating light fields are no longer pumped identically, i.e., where $\theta_1 \neq \theta_2$ and/or $P^\mathrm{in}_1 \neq P^\mathrm{in}_2$. 
This problem has been studied numerically in Ref.~\citenum{GarbinPRR2020}, and here we not only address it analytically, but also confirm and extend their results.
To remain consistent with the conventions established therein, we express Eq.~\eqref{eq:eqsintensitiesgeneral} using the substitutions $\theta_1 = \theta$, $\theta_2 = \theta - \Delta$, $P^\mathrm{in}_1 = P^\mathrm{in} \cos^2 \chi$, and $P^\mathrm{in}_2 = P^\mathrm{in} \sin^2 \chi$ as
\begin{align}\label{eq:intensitiesimbalanced}
	P^\mathrm{in}\sin^2\chi &- \left[ 1+ (-\theta + \Delta +  A P_2 + B P_1)^2\right]P_2 =0, \notag \\
    P^\mathrm{in}\cos^2\chi &- \left[ 1+ (-\theta +  A P_1 + B P_2)^2\right]P_1 =0.
\end{align}
Here, $\chi$ is the ellipticity angle, which captures the imbalance in pumping strength between the two fields, so that $P^\mathrm{in}=P_1^\mathrm{in}+P_2^\mathrm{in}$, while $\Delta$ represents the detuning difference between the two cavity modes.
By setting $\chi = \pi/4$ and $\Delta = 0$ we retrieve balanced input conditions up to a factor $1/2$ on the input intensity, cf. Eq.~\eqref{eq:intensitiesbalanced}.

Like for the balanced input parameters, univariate polynomials in one of the intensities can be constructed using the resultant method.
Namely, let $f_1(P_1, P_2)$ and $f_2(P_1,P_2)$ be the polynomial functions defined in Eq.~\eqref{eq:intensitiesimbalanced}. The resultants $\mathrm{res}(f_1, f_2, P_1)$ and $\mathrm{res}(f_1, f_2, P_2)$ are polynomials in the variables $P_2$ and $P_1$, respectively, whose coefficients depend on $\chi$ and~$\Delta$. The coefficients of these polynomials are presented in the SI, as well for the equations written in terms of the field amplitudes.
The roots of the resultants encode all solutions of Eq.~\eqref{eq:intensitiesimbalanced}.

To assess the bifurcations for imbalanced input parameters, we first realize that both resultants are generally different as the exchange symmetry between $P_1$ and $P_2$ is broken.
The bifurcations in the nonlinear system occur at the points in the parameter space, spanned by $\chi$ and $\Delta$, at which the discriminants of the resultant polynomials simultaneously vanish, namely, \mbox{$\mathrm{dis}\left(\mathrm{res}(f_1, f_2, P_2), P_1\right)=0$} and \mbox{$\mathrm{dis}\left(\mathrm{res}(f_1, f_2, P_1), P_2\right)=0$}.
Otherwise, crossing between the roots of the polynomials still occur, but this does not correspond to a global dynamical transition involving both fields.

Even under asymmetric input conditions, both intensities can become equal and satisfy the stationary state equations. This implies, consistently with Ref.~\citenum{GarbinPRR2020}, that the asymmetry introduced by the ellipticity $\chi$ can be compensated by an appropriate detuning difference $\Delta$, and vice versa. However, this compensation occurs for only one point on the possible solution branches, and full symmetry restoration across all stationary states is not possible. By imposing the condition $P_1 = P_2$ in Eq.~\eqref{eq:intensitiesimbalanced}, we can determine the resultant and find all parameters $\chi$ and $\Delta$ satisfying this condition.
These parameters define a symmetry-restoring curve in parameter space, but they only correspond to bifurcation points if the stricter requirement discussed before is met.

Figure~\ref{fig:fig5}(a) shows the parameter space defined by the ellipticity angle $\chi$ and the detuning difference $\Delta$, for fixed values of $P^\mathrm{in}$, and $\theta$. As shown in panels (b) to (d), the bifurcations correspond to the limits of optical bistability. The dot-dashed lines in panel (a) correspond to the points at which both intensities $P_1$ and $P_2$ are equal.

Importantly, even if the roots of the polynomials associated with $P_1$ and $P_2$ are equal, they may not simultaneously satisfy both stationary state equations defined in Eq.~\eqref{eq:intensitiesimbalanced}.
This defines a region at which asymmetric balance is fulfilled, as introduced in Ref.~\citenum{GarbinPRR2020}, cf. SI~\cite{SI}.

\section{Conclusions} \label{section:conclusions}
In this work, we present an analytical framework for obtaining polynomial equations that describe the homogeneous stationary states of circulating field intensities and amplitudes in Kerr ring and Fabry-Pérot resonators. By leveraging techniques from nonlinear algebra, namely polynomial resultants and Gröbner bases, we analytically captured the bifurcations underlying optical bistability and SSB.

Our findings not only clarify the structure and bifurcations of stationary states in Kerr resonators, but also contribute to the growing body of literature that connects nonlinear and non-Hermitian systems~\cite{KepesidisNJP2016, WangOL2019, SuntharalingamNatComms2023,WingenbachPRR2024, KwongarXiv2025, SuntharlingamPRAppl2025, WeisPRR2025, KawabataPRL2025}.
In case of balanced input conditions, we mapped our nonlinear system to a linear non-Hermitian auxiliary system, whose eigenvalues map back to the solutions of the nonlinear system.
Also, we identified the bifurcations describing optical bistability and SSB as EPs in the auxiliary system, similar to Refs.~\onlinecite{BaiNSR2023, BaiPRL2024,FangPRB2025, GuRiP2024}.
However, it should be emphasized that this mapping (i) is not unique, and (ii) the auxiliary non-Hermitian system implements a different physical system: all its eigenvalues correspond to physical solutions (no physicality constraint as in the nonlinear case), and all solutions are stable --
in the nonlinear case, the stability needs to be assessed via linear stability analysis~\cite{WoodleyPRA2018, BithaPRE2023, HillarXiv2025}.
The physical solutions we found, for both symmetric and asymmetric cases, may be unstable.

Beyond that, our results confirm and supplement those presented in Ref.~\citenum{GhoshLPR2025}, where it is shown that SSB occurs not only in the intensities, but also in the phases of the circulating fields within the resonators.
Here, we solved the amplitude equations, which contain the information for the intensities and the phases.
We also confirm and extend the results presented in Ref.~\citenum{GarbinPRR2020} by showing the full asymmetric balance regions and the symmetry restoration curves.
In Ref.~\citenum{HillarXiv2025}, an indirect connection between EPs, stemming from the linear stability analysis, and optical bistability as well as SSB has been made, and we provide a direct connection in this work.

The framework presented here provides a generalizable tool for studying Kerr-type nonlinearities with full analytical control.
Future work could explore how this framework extends to more general nonlinear coupled-mode equations as in Refs.~\citenum{GhoshPRR2023, PalPR2024, GhoshCLEO2025, YanArXiv2025}, and other types of nonlinearities.
Having full analytic insight enables better control for the engineering of bistable phenomena and spontaneous symmetry breaking for potential applications in the design of robust photonic circuits and all-optical logic devices.

\section{Methods}\label{section:Methods}

Our approach is based on nonlinear algebra and specifically elimination theory, namely, computing a resultant polynomial between the stationary state equations in terms of the intensities, or constructing Gröbner bases for the set of polynomials in terms of the field amplitudes.
Both methods can be applied for nonlinear systems with polynomial nonlinearities~\cite{Broer2003, hosakadoIEEE2001}.
Similar techniques have been employed to study diverse systems, including the Mandelbrot set~\cite{GeumChaos2009} and predator-prey population dynamics models~\cite{HajnovaMB2019}.

\subsection{Resultants}\label{subsection-Intensities}

The resultant polynomial of two polynomials $f(x) = \sum_{j=0} ^n \alpha_j x^{j} $ and $g(x)= \sum_{j=0} ^m \beta_j x^{j} $, with \mbox{$\alpha_n, \beta_m \neq 0$}, is an expression that depends only on the coefficients of $f$ and $g$. It is  computed as the determinant of the Sylvester matrix \cite{Cox2005}, namely
\begin{multline*}
   \mathrm{res}(f,g,x)=\\
   \begin{vmatrix} \begin{pmatrix}
\alpha_n & 0 & \cdots &  & \beta_m & 0 &  & 0 \\
\alpha_{n-1} & \alpha_n  &  &  & \beta_{m-1} & \beta_m &  & \vdots \\
\alpha_{n-2} & \alpha_{n-1} & \ddots &  & \beta_{m-2} & \beta_{m-1} &  &  \\
\vdots & \alpha_{n-2} & \ddots & \alpha_n & \vdots & \beta_{m-2} & \ddots &\beta_m  \\
\alpha_0 & \vdots & \ddots & \alpha_{n-1} & \beta_0 & \vdots & \ddots & \beta_{m-1} \\
0 & \alpha_0 &  &\alpha_{n-2}  & 0 & \beta_0 & \ddots & \beta_{m-2} \\
\vdots & 0 & \ddots & \vdots & \vdots & 0 & \ddots & \vdots \\
0 & 0 &  & \alpha_0 &  & \vdots &  & \beta_0
\end{pmatrix}\end{vmatrix}.
\end{multline*}
The resultant of two polynomials vanishes when both polynomials share a root, i.e., when simultaneously $f(x)=0$ and $g(x)=0$ for a specific $x$.
When the polynomials $f$ and $g$ are multivariate, this method allows eliminating one of the variables by taking the remaining variables as coefficients.

An illustrative example can be found in the SI.

\subsection{Gröbner Bases}

The second technique we employ are Gröbner bases.
It is a technique employed in elimination theory, which generalizes the algorithm of Gaussian elimination, and the algorithm for computing the greatest common divisor of univariate polynomials.
For our intends and purposes, it can be seen as the generalization of the resultant method for arbitrary many polynomials and variables.
The Gröbner basis $G$ of a set of polynomials $I$ is a subset of polynomials built from $I$ such that any element of $f \in I$ has a remainder zero when it is divided by any element $g \in G$ \cite{Cox2015}.
Depending on the lexicographic order, i.e., the order of the variables, different basis sets may be constructed.
The elimination theorem~\cite{Cox2015} ensures that a univariate polynomial, the object of interest in the main text, can be obtained leveraging Gröbner bases.

Again, an example on how to apply this method is given in the SI.

\subsection{ Discriminants and Bifurcations}

The final tool we introduce is the discriminant.
Defining the univariate polynomial $h(x) = \sum_{j=0}^k r_j x^j$, its discriminant $\mathrm{dis}$ is given by
\begin{equation}\label{eq:discriminant}
    \mathrm{dis}(h) = \frac{(-1)^{k(k-1)/2}}{r_k}\,\mathrm{res}\left(h, \frac{\dd h}{\dd x},x \right).
\end{equation}
The discriminant is a polynomial in the coefficients $r_j$, which vanishes if and only if $h(x)$ has a repeated root~\cite{Cox2015}.

In the main text, we encode the solutions of the nonlinear equations in univariate polynomials, and we identify points, at which the discriminant vanishes with different types of bifurcations.
In the matrix picture, such points correspond to degeneracies of $M$.

Furthermore, we left out some of the analytical expressions for bifurcations in the main text, which we present here.
For typographic reasons, we replace the superscripts of the input fields and intensities to subscripts, i.e., we use $E_\mathrm{in}$ and $P_\mathrm{in}$ instead of $E^\mathrm{in}$ and $P^\mathrm{in}$ in this section.
We also set $A=1$ and $B=2$ throughout this section, the full expressions for all polynomials can be found in the SI.

The first expression we present here is for balanced parameters in the intensity picture.
The discriminant of the polynomial $p^\mathrm{a}$ is given by
\begin{align}
&\mathrm{dis} \left(p^\mathrm{a}(P), P\right) = -331776 \theta ^6 P_\mathrm{in}^2 \notag \\
& \times \left(4 \theta ^4+8 \theta ^2+3 P_\mathrm{in}^2\right.
\left.-12 \theta ^3 P_\mathrm{in}+20 \theta  P_\mathrm{in}+4\right) \label{eq:dispa}\\
& \times\left(4 \theta ^4+72 \theta ^2+243 P_\mathrm{in}^2-4 \theta ^3 P_\mathrm{in}-324 \theta  P_\mathrm{in}+324\right)^2.\notag   
\end{align}
Analyzing this expression, we cannot only determine the bifurcations of $p^\mathrm{a}$, but also how many physical solutions there are.
As $p^\mathrm{a}$ is a real sixth-order polynomial, physical solutions can only appear in even numbers: zero solutions (no SSB), two (onset of SSB and having only SSB), four (onset of optical bistability of the asymmetric solutions) and six (optical bistability of the asymmetric solutions).
Because the sign of Eq.~\eqref{eq:dispa} only depends on the sign of the term in the second line of Eq.~\eqref{eq:dispa}, the expression in the second line determines when the polynomial $p^\mathrm{a}$ has two or zero real roots, and therefore determines if there is SSB or not, respectively.
If $\mathrm{dis} \left(p^\mathrm{a}(P), P\right) >0$, then $p^\mathrm{a}$ can have two or six different real roots: two if there is only SSB, and six if there is optical bistability of the asymmetric solution.
Thus, the factor in the third line determines the limits of optical bistability in the asymmetric regime, as shown in Fig.~\ref{fig:fig3}.

For balanced parameters in the amplitudes picture, we have set $\tilde{p}(E)=\tilde{p}^\mathrm{s}(E)\tilde{p}^\mathrm{a}(E)$, and the individual discriminants are
\begin{multline}\label{eq:distildeps}
    \mathrm{dis}(\tilde{p}^\mathrm{s}(E), E) = E_\mathrm{in}^2 \left(243 |E_\mathrm{in}|^4 -12 \theta ^3 |E_\mathrm{in}|^2\right.\\ \left.-108 \theta  |E_\mathrm{in}|^2
    +4 \theta ^4+8 \theta ^2+4\right),
\end{multline}
and
\begin{align}\label{eq:distildepa}
    &\mathrm{dis}(\tilde{p}^\mathrm{a}(E), E) = 64 E_\mathrm{in}^{10} |E_\mathrm{in}|^8 \notag \\
    & \times \left(4 \theta ^4+8 \theta ^2+ 3 |E_\mathrm{in}|^4 -12 \theta ^3 |E_\mathrm{in}|^2+20\theta  |E_\mathrm{in}|^2+4\right) \notag \\
    & \times \left(4 \theta ^4
   +72 \theta ^2+ 243 |E_\mathrm{in}|^4-4 \theta ^3|E_\mathrm{in}|^2-324 \theta |E_\mathrm{in}|^2
   \right.  \notag \\ 
   & \phantom{\times \big(} \left. +324\right)^2.
\end{align}
Here, it is important to notice that all factors in the discriminants of the symmetric (asymmetric) polynomials, Eqs.~\eqref{eq:disps} and \eqref{eq:distildeps} (Eqs.~\eqref{eq:dispa} and \eqref{eq:distildepa}), are identical up to their respective factors in the beginning.
These factors only vanish when  $P_\mathrm{in}=0$ or $E_\mathrm{in}=0$ or $\theta =0$, not representing physical bifurcations.
Thus, this shows that the bifurcations appear in the same points in parameter space in the intensity and amplitude pictures.
Furthermore, it is now also apparent from Eqs.~\eqref{eq:distildeps} and \eqref{eq:distildepa} that the only input-phase-dependent term $E_\mathrm{in}$ drops out when determining the bifurcations for the amplitudes.

\section*{Data Availability}
The authors declare that the data supporting the findings of this study are available within the paper, and its supplementary information file.

\section*{Author Contributions}
J.D.M.-V. carried out the calculations and theoretical work. J.T.G., L.H., and F.K.K. contributed to discussions, the analysis of the results, and provided ongoing guidance.
J.T.G. initiated and, together with L.H., shaped the project. J.D.M.-V. and J.T.G prepared the manuscript with support of L. H. and input from F.K.K.

\section*{Acknowledgments}

J.D.M.-V. thanks Nicolás Salcedo-Gálvez for his help with Fig.~\ref{fig:fig1}. We acknowledge support from the MPG Lise Meitner Excellence Program 2.0. J.D.M.-V. and F.K.K. are part of the Max Planck School of Photonics supported by the Dieter Schwarz Foundation, the German Federal Ministry of Research, Technology and Space (BMFTR), and the Max Planck Society. We also acknowledge funding from the European Union's ERC Starting Grant “NTopQuant” (101116680). The views expressed are those of the authors and do not necessarily reflect those of the EU or the ERC. Neither the EU nor the granting authority can be held responsible for them.

\bibliography{referencesNOChange}

\onecolumngrid

\clearpage

\setcounter{figure}{0}
\renewcommand{\thefigure}{S\arabic{figure}}
\renewcommand{\theequation}{S\arabic{equation}}
\setcounter{equation}{0}
\setcounter{section}{0}
\setcounter{table}{0}

\allowdisplaybreaks

\onecolumngrid
\part*{\large\centering Supplementary Information:\\ An Algebraic Approach to Bifurcations in Kerr Ring and Fabry-Pérot Resonators}
\begin{center}
    J. D. Mazo-Vásquez, J. T. Gohsrich, F. K. Kunst, and L. Hill
\end{center}

\section{List of Polynomial Coefficients}
In this section, we include the coefficients of the polynomials obtained by either using the resultant method (for the analysis of the field intensities) or Gröbner bases (for the analysis of the field amplitudes). We replace here the superscripts of the input fields and intensities to subscripts, i.e., we use $E_\mathrm{in}$ and $P_\mathrm{in}$ instead of $E^\mathrm{in}$ and $P^\mathrm{in}$.

\subsection{Imbalanced Input Parameters: Intensities}\label{sec:intensity-imbalanced}
The resultants of \mbox{Eq.~\eqref{eq:intensitiesimbalanced}} in the main text with respect to the variables $P_1$ and $P_2$ lead to a ninth-order polynomial in $P_2$, and $P_1$, respectively. When $P_2$ is eliminated, the resultant is the polynomial $p(P_1) = \sum_{j=0}^9 a_j P_1^j$, where the coefficients are given by
\begin{align*}
    a_0&=A^4 P_\mathrm{in}^3 \cos ^6(\chi ),\\
    a_1&=-A^2 P_\mathrm{in}^2 \cos ^4(\chi ) (3 A^2 (\theta ^2+1)+4 A B \theta  (\Delta -\theta )+2 B^2 ((\Delta -\theta )^2-1)),\\
    a_2&=P_\mathrm{in} \cos ^2(\chi ) (3 A^5 \theta  P_\mathrm{in}+A^4 (2 B P_\mathrm{in} (\Delta -\theta )+3 (\theta ^2+1)^2)-2 A^3 B \theta  (B P_\mathrm{in}-4 (\theta ^2+1) (\Delta -\theta ))+A^2 B^2 (B P_\mathrm{in} (5 \theta \\ &
    -2 \Delta )+4 (2 \theta ^2+1) (\Delta -\theta )^2-4)+A P_\mathrm{in} \cos (2 \chi ) (3 A^4 \theta +2 A^3 B (\Delta -\theta )-2 A^2 B^2 \theta -A B^3 (2 \Delta +\theta )+2 B^4 (\theta -\Delta ))\\
    &+2 A B^3 (\Delta -\theta ) (B P_\mathrm{in}+2 \theta  ((\Delta -\theta )^2+1))+B^4 ((\Delta -\theta )^2+1)^2),\\
    a_3& =-((\theta ^2+1) (A^2 (\theta ^2+1)+2 A B (\theta  (\Delta -\theta )-1)+B^2 ((\Delta -\theta )^2+1)) (A^2 (\theta ^2+1)+2 A B (\theta  (\Delta -\theta )+1)\\&+B^2 ((\Delta -\theta )^2+1)))-P_\mathrm{in} (-2 B^3 \sin ^2(\chi ) (A^2 \theta  (\theta ^2-3)+2 A B (\theta ^2-1) (\Delta -\theta )+B^2 \theta  ((\Delta -\theta )^2+1))\\
    &+A^2 P_\mathrm{in} (3 A^4-4 A^2 B^2+2 B^4) \cos ^4(\chi )+\cos ^2(\chi ) (12 A^5 (\theta ^3+\theta )+8 A^4 B (3 \theta ^2+1) (\Delta -\theta )-3 A^3 B^3 P_\mathrm{in} \cos (2 \chi )\\
    &+A^3 B^2 (3 B P_\mathrm{in}+8 \theta  (2 \Delta ^2-4 \Delta  \theta +\theta ^2-1))+4 A^2 B^3 (\Delta -\theta ) ((\Delta -3 \theta ) (\Delta +\theta )-1)-4 A B^4 \theta  (3 (\Delta -\theta )^2+1)\\
    &-4 B^5 ((\Delta -\theta )^2+1) (\Delta -\theta ))-A B^5 P_\mathrm{in} \sin ^2(2 \chi )+B^6 P_\mathrm{in} \sin ^4(\chi )),\\
    a_4 &=2 (3 A^5 \theta  (\theta ^2+1)^2+2 A^4 B (5 \theta ^4+6 \theta ^2+1) (\Delta -\theta )+2 A^3 B^2 \theta  (\Delta ^2 (6 \theta ^2+4)-4 \Delta  (3 \theta ^2+2) \theta \\
    &+5 \theta ^4+4 \theta ^2-1)+B^3 P_\mathrm{in} \sin ^2(\chi ) (3 A^3-A B^2 (\Delta ^2+3)-2 B \Delta  \theta  (A-B) (2 A+B)-\theta ^2 (A-B)^2 (3 A+2 B))\\
    &+2 A^2 B^3 \Delta  (3 \theta ^2+1) (\Delta -\theta ) (\Delta -2 \theta )+P_\mathrm{in} \cos ^2(\chi ) (A^6 (9 \theta ^2+3)+12 A^5 B \theta  (\Delta -\theta )\\
    &+4 A^4 B^2 (\Delta ^2-2 \Delta  \theta -2 \theta ^2-1)+16 A^3 B^3 \theta  (\theta -\Delta )-2 A^2 B^4 (3 \Delta ^2-6 \Delta  \theta +\theta ^2)+6 A B^5 \theta  (\Delta -\theta )\\
    &+B^6 (3 (\Delta -\theta )^2+1))+A B^4 \theta  (\Delta ^4-4 \Delta ^3 \theta -4 \Delta ^2+8 \Delta  (\theta ^3+\theta )-(\theta ^2+1) (5 \theta ^2+1))\\
    &-2 B^5 (\theta ^2+1) ((\Delta -\theta )^2+1) (\Delta -\theta )),\\
    a_5& =-3 A^6 (5 \theta ^4+6 \theta ^2+1)-8 A^5 B \theta  (5 \theta ^2+3) (\Delta -\theta )+4 A^4 B^2 (-\Delta ^2 (9 \theta ^2+2)+2 \Delta  (9 \theta ^2+2) \theta\\
    &-4 \theta ^4+\theta ^2+1)-4 A^3 B^3 \theta  (\Delta -\theta ) (3 (\Delta -3 \theta ) (\Delta +\theta )-5)+A^2 B^4 (-\Delta ^4+2 (15 \Delta ^2+7) \theta ^2\\
    &+4 (\Delta ^2-5) \Delta  \theta +10 \Delta ^2-68 \Delta  \theta ^3+29 \theta ^4+1)+2 P_\mathrm{in} (A-B) (A+B) (B^3 \sin ^2(\chi ) (3 A^2 \theta +2 A B (\Delta -\theta )\\
    &-B^2 \theta )-2 \cos ^2(\chi ) (B \Delta  (2 A^4-2 A^2 B^2+B^4)+\theta  (A-B) (3 A^4+A^3 B-2 A^2 B^2+B^4)))+4 A B^5 \theta  (\Delta -\theta ) (2 \Delta ^2\\
    &-4 \Delta  \theta -\theta ^2-1)-2 B^6 (\theta ^2+1) (3 (\Delta -\theta )^2+1),\\
    a_6&=\frac{1}{2} (A-B) (A+B) (3 A^6 P_\mathrm{in}+8 A^5 \theta  (5 \theta ^2+3)+A^4 B (16 (5 \theta ^2+1) (\Delta -\theta )-5 B P_\mathrm{in})-2 A^3 B^2 (B P_\mathrm{in}\\
    &-4 \theta  (6 \Delta ^2-12 \Delta  \theta +\theta ^2-1))+A^2 B^3 (3 B P_\mathrm{in}+8 (\Delta -\theta ) (\Delta ^2-2 \Delta  \theta -7 \theta ^2-1))\\
    &+P_\mathrm{in} (A+B)^2 (3 A^4-6 A^3 B+4 A^2 B^2-B^4) \cos (2 \chi )+2 A B^4 (B P_\mathrm{in}-4 \theta  (3 \Delta ^2-6 \Delta  \theta +2 \theta ^2))\\
    &-B^5 (B P_\mathrm{in}-8 (\theta ^2+1) (\Delta -\theta ))), \\
    a_7&=-(A-B)^2 (A+B)^2 (3 A^4+6 A^2 B^2 \Delta ^2+\theta ^2 (A-B)^2 (15 A^2+10 A B+B^2)+4 A B \Delta  \theta  (A-B) (5 A+2 B)+B^4),\\
    a_8&=2 A (A^2-B^2)^3 (3 A^2 \theta +2 A B (\Delta -\theta )-B^2 \theta ),\\
    a_9&=-A^2 (A^2-B^2)^4.
\end{align*}
If $P_1$ is eliminated, the resultant polynomial is $p(P_2) = \sum_{j=0}^9 b_j P_2^j$, where the coefficients are
\begin{align*}
b_0 & = A^4 P_\mathrm{in}^3 \sin ^6(\chi ), \\
b_1 &=  -A^2 P_\mathrm{in}^2 \sin ^4(\chi ) (3 A^2 ((\Delta -\theta )^2+1)+4 A B \theta  (\Delta -\theta )+2 B^2 (\theta ^2-1)), \\
b_2 &= P_\mathrm{in} \sin ^2(\chi ) (3 A^5 P_\mathrm{in} (\theta -\Delta )+A^4 (-2 B \theta  P_\mathrm{in}+3 ((\Delta -\theta )^2+2) (\Delta -\theta )^2+3)+2 A^3 B (\Delta -\theta ) (B P_\mathrm{in}\\
&+4 \theta  ((\Delta -\theta )^2+1))+A^2 B^2 (B P_\mathrm{in} (5 \theta -3 \Delta )+4 \theta ^2 (2 (\Delta -\theta )^2+1)-4)+A P_\mathrm{in} \cos (2 \chi ) (3 A^4 (\Delta -\theta)\\
&+2 A^3 B \theta +2 A^2 B^2 (\theta -\Delta )+A B^3 (\theta -3 \Delta )-2 B^4 \theta )-2 A B^3 \theta  (B P_\mathrm{in}-2 (\theta ^2+1) (\Delta -\theta ))+B^4 (\theta ^2+1)^2),\\
b_3 &= -((\Delta -\theta )^2+1) (A^2 ((\Delta -\theta )^2+1)+2 A B (\theta  (\Delta -\theta )-1)+B^2 (\theta ^2+1)) (A^2 ((\Delta -\theta )^2+1)\\
&+2 A B (\theta  (\Delta -\theta )+1)
+B^2 (\theta ^2+1))-A^2 P_\mathrm{in}^2 (3 A^4-4 A^2 B^2+2 B^4) \sin ^4(\chi )-B^3 P_\mathrm{in} \cos ^2(\chi )\\
&\times(-3 A^3 P_\mathrm{in} \cos (2 \chi )+3 A^3 P_\mathrm{in}
+2 A^2 ((\Delta -\theta )^2-3) (\Delta -\theta )+4 A B \theta  ((\Delta -\theta )^2-1)+2 B^2 (\theta ^2+1) (\Delta -\theta ))\\
&+4 P_\mathrm{in} \sin ^2(\chi )(3 A^5 ((\Delta -\theta )^2+1) (\Delta -\theta )+2 A^4 B \theta  (3 (\Delta -\theta )^2+1)-2 A^3 B^2 (\Delta -\theta ) (\Delta ^2-2 \Delta  \theta -\theta ^2+1)\\
&-A^2 B^3 \theta  (4 \Delta ^2-8 \Delta  \theta +3 \theta ^2+1)-A B^4 (3 \theta ^2+1) (\Delta -\theta )-B^5 \theta  (\theta ^2+1))+A B^5 P_\mathrm{in}^2 \sin ^2(2 \chi )\\
&+B^6 (-P_\mathrm{in}^2) \cos ^4(\chi ),\\
b_4 &= 2 (-3 A^5 ((\Delta -\theta )^2+1)^2 (\Delta -\theta )-2 A^4 B \theta  (5 (\Delta -\theta )^4+6 (\Delta -\theta )^2+1)+2 A^3 B^2 (\Delta -\theta ) (\Delta ^4\\
&-4 \Delta ^3 \theta +8 \Delta  \theta ^3-5 \theta ^4-4 \theta ^2+1)+2 A^2 B^3 \Delta  \theta  (3 (\Delta -\theta )^2+1) (\Delta -2 \theta )+B^3 P_\mathrm{in} \cos ^2(\chi )\\
&\times(-3 A^3 ((\Delta -\theta )^2-1)+4 A^2 B \theta  (\theta -\Delta )+A B^2 (2 \Delta ^2-4 \Delta  \theta +\theta ^2-3)+2 B^3 \theta  (\Delta -\theta ))+P_\mathrm{in} \sin ^2(\chi )\\
&\times(3 A^6 (3 (\Delta -\theta )^2+1)+12 A^5 B \theta  (\Delta -\theta)-4 A^4 B^2 (3 \Delta ^2-6 \Delta  \theta +2 \theta ^2+1)+16 A^3 B^3 \theta  (\theta -\Delta )\\
&+2 A^2 B^4 (2 \Delta ^2-4 \Delta  \theta -\theta ^2)+6 A B^5 \theta  (\Delta -\theta )+B^6 (3 \theta ^2+1))+A B^4 (\Delta -\theta ) (\Delta ^2 (6 \theta ^2+2)-4 \Delta  (3 \theta ^3+\theta )\\
&+5 \theta ^4+6 \theta ^2+1)+2 B^5 \theta  (\theta ^2+1) ((\Delta -\theta )^2+1)),\\
b_5&= -3 A^6 (5 (\Delta -\theta )^4+6 (\Delta -\theta )^2+1)-8 A^5 B \theta  (5 (\Delta -\theta )^2+3) (\Delta -\theta )+4 A^4 B^2 (5 \Delta ^4-20 \Delta ^3 \theta\\
&+3 \Delta ^2 (7 \theta ^2+1)-2 \Delta  \theta  (\theta ^2+3)-4 \theta ^4+\theta ^2+1)+4 A^3 B^3 \theta  (\Delta -\theta ) (12 \Delta ^2-24 \Delta  \theta +9 \theta ^2+5)\\
&+A^2 B^4 (-6 \Delta ^4+24 \Delta ^3 \theta +4 \Delta ^2-8 \Delta  (6 \theta ^3+\theta )+29 \theta ^4+14 \theta ^2+1)+2 P_\mathrm{in} (A-B) (A+B) (B^3 \cos ^2(\chi )\\
&\times(3 A^2 (\theta -\Delta )-2 A B \theta +B^2 (\Delta -\theta ))+2 \sin ^2(\chi ) (A \Delta  (3 A^4-3 A^2 B^2+B^4)-\theta  (A-B) (3 A^4\\
&+A^3 B-2 A^2 B^2+B^4)))-4 A B^5 \theta  (\Delta -\theta ) (3 \Delta ^2-6 \Delta  \theta +\theta ^2+1)-2 B^6 (3 \theta ^2+1) ((\Delta -\theta )^2+1),\\
b_6&= \frac{1}{2} (A-B) (A+B) (3 A^6 P_\mathrm{in}-8 A^5 (5 (\Delta -\theta )^2+3) (\Delta -\theta )-A^4 B (5 B P_\mathrm{in}+16 \theta  (5 (\Delta -\theta )^2+1))\\
&+2 A^3 B^2 (4 (\Delta -\theta ) (5 \Delta ^2-10 \Delta  \theta -\theta ^2+1)-B P_\mathrm{in})+A^2 B^3 (3 B P_\mathrm{in}+8 \theta  (8 \Delta ^2-16 \Delta  \theta +7 \theta ^2+1))\\
&-P_\mathrm{in} (A+B)^2 (3 A^4-6 A^3 B+4 A^2 B^2-B^4) \cos (2 \chi )+2 A B^4 (B P_\mathrm{in}-4 \Delta ^3+12 \Delta ^2 \theta -8 \theta ^3)\\
&-B^5 (B P_\mathrm{in}+8 \theta  ((\Delta -\theta )^2+1))),\\
b_7&= -(A-B)^2 (A+B)^2 (3 A^4 (5 (\Delta -\theta )^2+1)+20 A^3 B \theta  (\Delta -\theta )-2 A^2 B^2 (5 \Delta ^2-10 \Delta  \theta +2 \theta ^2)\\
&+8 A B^3 \theta  (\theta -\Delta )+B^4 ((\Delta -\theta )^2+1)), \\
b_8&= -2 A (A^2-B^2)^3 (3 A^2 (\Delta -\theta )+2 A B \theta +B^2 (\theta -\Delta )), \\
b_9&= - A^2 (A^2 - B^2)^4.
\end{align*}

\subsection{Balanced Input Parameters: Intensities}\label{sec:intensity-balanced}
If balanced input conditions are considered, i.e., $\chi= \pi/4$ and $\Delta=0$, the polynomials $p(P_2)$ and $p(P_1)$ become equal to $p(P) = p^\mathrm{s}(P)p^\mathrm{a}(P)$, as in Eq.~\eqref{eq:polsintensities} in the main text, up to a factor of $1/2$ for $P_\mathrm{in}$.
The polynomial $p^\mathrm{s}(P)$ is given in Eq.~\eqref{eq:symmetricintensities}, and the coefficients of the polynomial $p^\mathrm{a}(P) = \sum_{j=0}^6 \gamma_j P^j$ are
\begin{align*}
    \gamma_6 &= -\frac{1}{2} A^2 (A-B)^4 (A+B)^2,\\
    \gamma_5 &=  A \theta  (A-B)^4 (2 A^2+3 A B+B^2),\\
    \gamma_4 &= -\frac{1}{2} (A-B)^2 (A^4 (6 \theta ^2+2)+A^3 B (2-6 \theta ^2)-A^2 B^2 (5 \theta ^2+1)+4 A B^3 \theta ^2+B^4 (\theta ^2+1)),\\
    \gamma_3 &= \frac{1}{4}(A-B)^2 (2 A^4 P_\mathrm{in}+2 A^3 (B P_\mathrm{in}+4 (\theta ^3+\theta ))+A^2 B (-3 B P_\mathrm{in}-12 \theta ^3+4 \theta )-2 A B^3 P_\mathrm{in}+B^3 (B P_\mathrm{in}+4 (\theta ^3+\theta ))),\\
    \gamma_2 &= -\frac{1}{2}(A-B)^2 (2 A^3 \theta  P_\mathrm{in}+A^2 (B \theta  P_\mathrm{in}+(\theta ^2+1)^2)-2 A B (B \theta  P_\mathrm{in}+\theta ^4-1)+B^2 (\theta ^2+1)^2),\\
    \gamma_1 &= \frac{1}{2} A^2 P_\mathrm{in} (A-B) (A \theta ^2+A-B \theta ^2+B).
\end{align*}

\subsection{Imbalanced Input Parameters: Amplitudes}\label{app:amplitudes-unbalaced}
From the Gröber basis of polynomials in Eq.~\eqref{eq:fields-auxiliaries} with $E^\mathrm{in}_1 = E_\mathrm{in} \sin\left(\chi\right)$, $E^\mathrm{in}_2 = E_\mathrm{in} \cos\left(\chi\right)$, $\theta_1=\theta$, and $\theta_2=\theta-\Delta$, a univariate polynomial in $E_1$ is found to be $ \tilde{p}(E_1)=\sum_{j=0}^9 c_j E_1^j $, where the coefficients are
\begin{align*}
c_0 &= -\ii A^2 E_\mathrm{in}^6 \sin ^6(\chi ),\hspace{13.3cm}\\
c_1 &= A E_\mathrm{in}^5 \sin ^5(\chi ) (2 B (\theta -\Delta )-3 A (\theta -3 \ii)),\\
c_2 &= \ii E_\mathrm{in}^4 \sin ^4(\chi ) (3 A^2 (-11+\theta  (\theta -8 \ii))+4 A B (\theta -4 \ii) (\Delta -\theta )+B^2 ((\Delta -\theta )^2+1)), \\
c_3 &=E_\mathrm{in}^3 \sin ^3(\chi ) (-|E_\mathrm{in}|^2 (A (3 A^2-2 B^2) \sin ^2(\chi )+B^3 \cos ^2(\chi ))+A^2 (\theta -3 \ii) (-21+\theta  (\theta -18 \ii))\\
&+2 A B (-25+\theta  (\theta -14 \ii)) (\Delta -\theta )+B^2 (\theta -7 \ii) ((\Delta -\theta )^2+1)), \\
c_4 &= E_\mathrm{in}^2 \sin ^2(\chi ) (-6 A^2 (\theta -\ii) (-11+\theta  (\theta -8 \ii))-\ii |E_\mathrm{in}|^2 (-3 A^3 (\theta -3 \ii)+2 A^2 B (\theta -\Delta )\\
& +\cos (2 \chi ) (3 A^3 (\theta -3 \ii)+2 A^2 B (\Delta -\theta )-2 A B^2 (\theta -3 \ii)+B^3 (-\Delta +\theta +3 \ii))+2 A B^2 (\theta -3 \ii)\\
&+B^3 (\Delta -\theta +3 \ii))-4 A B (-19+3 \theta  (\theta -6 \ii)) (\Delta -\theta )-6 B^2 (\theta -3 \ii) ((\Delta -\theta )^2+1)), \\
c_5 &= E_\mathrm{in} \sin (\chi ) (4 (3 A^2 (\theta -3 \ii) (\theta -\ii)^2+2 A B (\theta -\ii) (3 \theta -7 \ii) (\Delta -\theta )\\
&+B^2 (3 \theta -5 \ii) ((\Delta -\theta )^2+1))+|E_\mathrm{in}|^2 (-12 B^3 \cos ^2(\chi )+\sin ^2(\chi ) (3 A^3 (-13+\theta  (\theta -10 \ii))\\
&+4 A^2 B (\theta -5 \ii) (\Delta -\theta )+A B^2 (\Delta ^2-2 \Delta  \theta -\theta  (\theta -20 \ii)+27)-2 B^3 (\theta -5 \ii) (\Delta -\theta )))), \\
c_6 &= \ii (3 A^4-4 A^2 B^2+B^4) | E_\mathrm{in}| ^4 \sin ^4(\chi )+4 |E_\mathrm{in}|^2 (2 B^3 \cos ^2(\chi )\\
&+\sin ^2(\chi ) (-3 A^3 (-3+\theta  (\theta -4 \ii))-4 A^2 B (\theta -2 \ii) (\Delta -\theta )+A B^2 (-\Delta ^2\\
&+2 (\Delta -4 \ii) \theta +\theta ^2-7)+2 B^3 (\theta -2 \ii) (\Delta -\theta )))-8 (\theta -\ii) (B^2+(B (\Delta -\theta )+A (\theta -\ii))^2), \\
c_7 &= E_\mathrm{in}^* \sin (\chi ) (|E_\mathrm{in}|^2 (A-B) (A+B)\sin ^2(\chi ) (3 A^2 (\theta -3 \ii)+2 A B (\Delta -\theta )\\
&-B^2 (\theta -3 \ii))+4 (3 A^3 (\theta -\ii)^2+4 A^2 B (\theta -\ii) (\Delta -\theta )+A B^2 (\Delta ^2-2 \Delta  \theta -\theta  (\theta -4 \ii)+3)\\
&-2 B^3 (\theta -\ii) (\Delta -\theta ))), \\
c_8 &= -2 (A-B) (A+B) (E_\mathrm{in}^*)^2 \sin ^2(\chi ) (3 A^2 (\theta -\ii)+2 A B (\Delta -\theta )-B^2 (\theta -\ii)), \\
c_9 &= A (A^2-B^2)^2(E_\mathrm{in}^*)^3 \sin ^3(\chi ).
\end{align*}

By eliminating $E_1$ in Eq.~\eqref{eq:fields-auxiliaries} in the main text, a polynomial on the variable $E_2$ can be found. This is $\tilde{p}(E_2) = \sum_{j=0}^9 d_j E_2^j$ where the coefficients are given by
   \begin{align*}
d_0 &= -\ii A^2 E_\mathrm{in}^6 \cos ^6(\chi ),\\
d_1 &= A E_\mathrm{in}^5 \cos ^5(\chi ) (2 B \theta +3 A (\Delta -\theta +3 \ii)),\\
d_2 &= \ii E_\mathrm{in}^4 \cos ^4(\chi ) (3 A^2 (\Delta ^2-2 (\Delta +4 \ii) \theta +8 \ii \Delta +\theta ^2-11)\\
&+4 A B \theta  (\Delta -\theta +4 \ii)+B^2 (\theta ^2+1)), \\
d_3 &=-E_\mathrm{in}^3 \cos ^3(\chi ) (|E_\mathrm{in}|^2 (A (3 A^2-2 B^2) \cos ^2(\chi )+B^3 \sin ^2(\chi ))+A^2 (\Delta -\theta +3 \ii) (\Delta ^2-2 (\Delta +9 \ii) \theta \\
&+18 \ii \Delta +\theta ^2-21)+2 A B \theta  (\Delta ^2-2 (\Delta +7 \ii) \theta +14 \ii \Delta +\theta ^2-25)+B^2 (\theta ^2+1) (\Delta -\theta +7 \ii)),\\
d_4 &=2 E_\mathrm{in}^2 \cos ^2(\chi ) (3 A^2 (\Delta -\theta +\ii) (\Delta ^2-2 (\Delta +4 \ii) \theta +8 \ii \Delta +\theta ^2-11)\\
&+E_\mathrm{in}^* (3 B^3 E_\mathrm{in} \sin ^2(\chi )-\ii E_\mathrm{in} \cos ^2(\chi ) (3 A^3 (\Delta -\theta +3 \ii)+2 A^2 B \theta -2 A B^2 (\Delta -\theta +3 \ii)\\
&-B^3 \theta ))+2 A B \theta  (3 \Delta ^2-6 \Delta  (\theta -3 \ii)+3 \theta  (\theta -6 \ii)-19)+3 B^2 (\theta ^2+1) (\Delta -\theta +3 \ii)), \\
d_5 &=E_\mathrm{in} \cos (\chi ) (|E_\mathrm{in}|^2 (-12 B^3 \sin ^2(\chi )+\cos ^2(\chi ) (3 A^3 (-13+\Delta  (\Delta +10 \ii))\\
&-2 (\Delta +5 \ii) \theta  (A-B) (3 A^2+A B-B^2)+A B^2 (27-2 \Delta  (\Delta +10 \ii))+\theta ^2 (A-B)^2 (3 A+2 B)))\\
&-4 (3 A^2 (\Delta -\theta +\ii)^2 (\Delta -\theta +3 \ii)+2 A B \theta  (\Delta -\theta +\ii) (3 \Delta -3 \theta +7 \ii)+B^2 (\theta ^2+1) (3 \Delta -3 \theta +5 \ii))), \\
d_6 &= \ii (3 A^4-4 A^2 B^2+B^4) |E_\mathrm{in}|^4 \cos ^4(\chi )-4 |E_\mathrm{in}|^2 (-2 B^3 \sin ^2(\chi )\\
&+\cos ^2(\chi ) (3 A^3 (\Delta +\ii) (\Delta +3 \ii)-2 (\Delta +2 \ii) \theta  (A-B) (3 A^2+A B-B^2)\\
&+A B^2 (7-2 \Delta  (\Delta +4 \ii))+\theta ^2 (A-B)^2 (3 A+2 B)))+8 (\Delta -\theta +\ii) (B^2+(B \theta +A (\Delta -\theta +\ii))^2), \\
d_7 &=E_\mathrm{in}^* \cos (\chi ) (4 (3 A^3 (\Delta +\ii)^2-2 (\Delta +\ii) \theta  (A-B) (3 A^2+A B-B^2)+A B^2 (3-2 \Delta  (\Delta+2 \ii))\\
&+\theta ^2 (A-B)^2 (3 A+2 B))-E_\mathrm{in} (A-B) (A+B) E_\mathrm{in}^* \cos ^2(\chi ) (3 A^2 (\Delta -\theta +3 \ii)+2 A B \theta +B^2 (-\Delta +\theta -3 \ii))),\\
d_8 &= 2 (A-B) (A+B) (E_\mathrm{in}^*)^2 \cos ^2(\chi ) (3 A^2 (\Delta -\theta +\ii)+2 A B \theta +B^2 (-\Delta +\theta -\ii)), \\
d_9 &= A (A^2-B^2)^2 (E_\mathrm{in}^*)^3 \cos ^3(\chi ).
\end{align*}

\subsection{Balanced Input Parameters: Amplitudes}\label{sec:amplitudes-balanced}
Under symmetric input conditions, i.e., $\chi=\pi/4$ and $\Delta=0$, $\tilde{p}(E_1)$ and $\tilde{p}(E_2)$ become equal, i.e. $\tilde{p}(E)$, where $E$ is either $E_1$ or $E_2$. In such a case, the polynomial $\tilde{p}(E)$ can be written as the product of two polynomials, $\tilde{p}(E)$ and $\tilde{p}(E)$. The coefficients of the polynomials $\tilde{p}^\mathrm{a}(E) = \sum_{j=0}^6 \rho_j E^j$ and $\tilde{p}^\mathrm{s}(E) = \sum_{j=0}^3 \eta_j E^j$ are given by:

\begin{align*}
    \rho_6 &= 2 A (A-B)^2 (A+B) (E_\mathrm{in}^*)^2,\\
    \rho_5 &= -2 \sqrt{2} (A-B) E_\mathrm{in}^* (2 A^2 (\theta -\ii)-A B (\theta +\ii)-B^2 (\theta -\ii)),\\
    \rho_4 &= 2 (A-B) (|E_\mathrm{in}|^2 (2 A^2 (\theta -3 \ii)-A B (\theta +3 \ii)-B^2 (\theta -3 \ii))+2 A (\theta -\ii)^2-2 B (\theta ^2+1)),\\
    \rho_3 &= 2 \sqrt{2} E_\mathrm{in} (6 A^2+\ii E_\mathrm{in} (A-B) (2 A-B) (A+B) E_\mathrm{in}^*-2 \theta ^2 (A-B)^2+8 \ii A \theta  (A-B)-2 B^2),\\
    \rho_2 &=2 E_\mathrm{in}^2 (-13 A^2+\theta ^2 (A-B)^2-10 \ii A \theta  (A-B)+B^2),\\
    \rho_1 &= 4 \sqrt{2} A E_\mathrm{in}^3 (3 A+\ii \theta  (A-B)),\\
    \rho_0 &= -4 A^2 E_\mathrm{in}^4,\\
    \eta_3 &= \sqrt{2} (A+B) E_\mathrm{in}^*,\\
    \eta_2 &= -2 \theta +2 \ii,\\
    \eta_1 &= \sqrt{2} E_\mathrm{in} (\theta -3 \ii),\\
    \eta_0 &= 2 \ii E_\mathrm{in}^2.
\end{align*}

\section{Regions of asymmetric balance}\label{sec:comparison}
\begin{figure}[hbt!]
    \centering
    \includegraphics[width=0.6\linewidth]{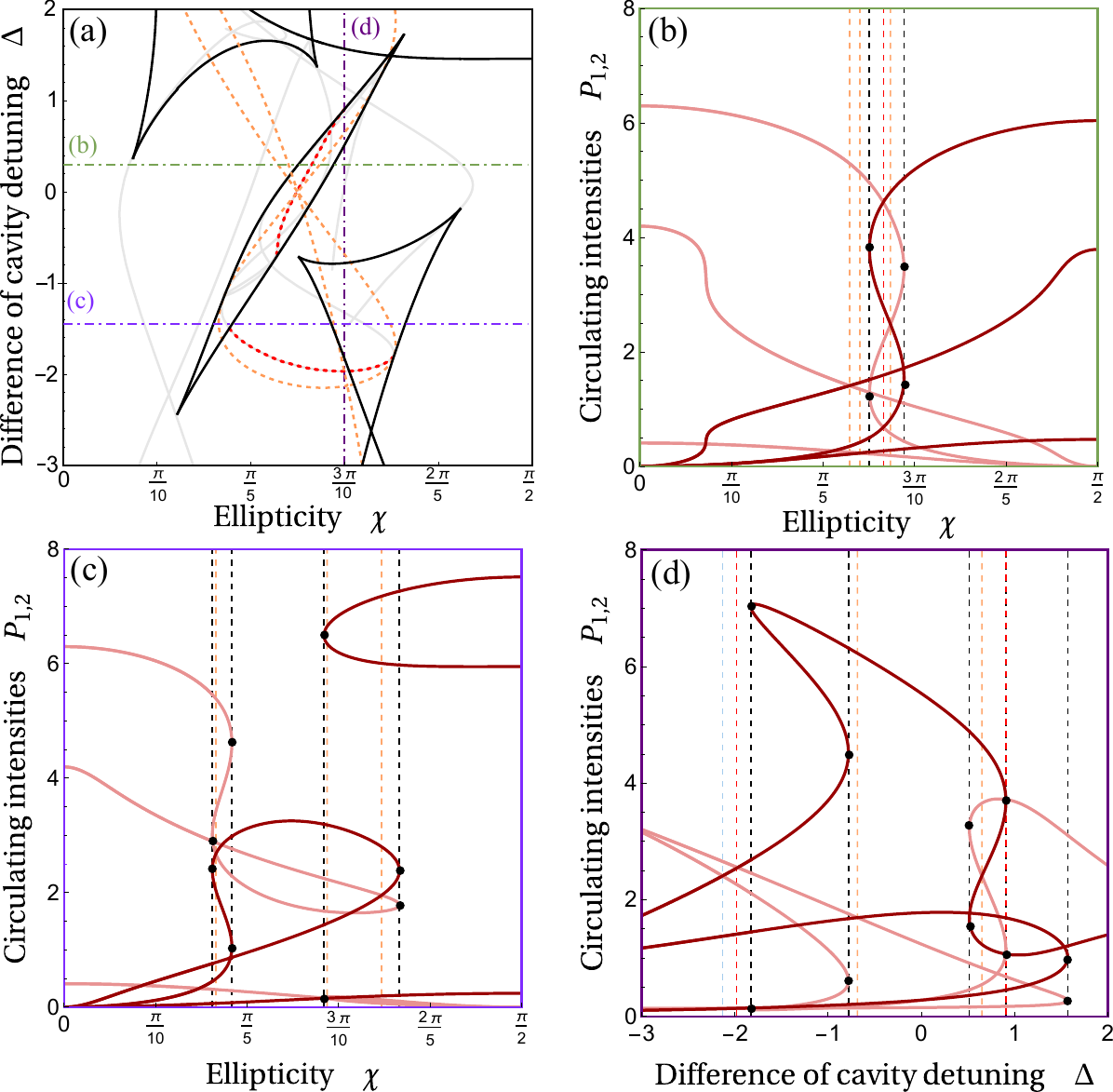}
    \caption{(a) Parameter space defined by $(\chi\,, \Delta)$ where bifurcations occur (black solid lines). Dot-dashed lines correspond to the parameter scans in (b-d). Circulating intensities as functions of the ellipticity $\chi$ in (b) and (c) for $\Delta = 0.3$, and $\Delta = -1.4$, respectively, and in (d) as functions of the difference of cavity detuning for $\chi = 3\pi/10$. The dashed black, orange and red lines in panels (b), (c), and (d) correspond to the position of the bifurcations, symmetric solutions for both circulating intensities, and the asymmetric balance lines in (a), respectively. The results shown here were computed with $A =1$, $B=1.57$, $P^\mathrm{in} =10.8$ and $\theta =5.45$. }
    \label{fig:figAppendix}
\end{figure}

We compare our analytical results by using the polynomials from the Gröbner basis for the set of polynomials in Eq.~\eqref{eq:intensitiesimbalanced} in the main text, whose coefficients are given in Sec.~\ref{sec:intensity-imbalanced}, with the results shown in Fig. 7(b) in Ref.~\cite{GarbinPRR2020}. The bifurcations are shown in Fig.~\ref{fig:figAppendix}(a), indicated by black solid lines. The dot-dashed lines indicate the paths along which the scans shown in panels (b-d) were determined. Panels (b,c) correspond to scans along the ellipticity angle $\chi$ for $\Delta = 0.3$ and $\Delta = -1.4$, respectively, whereas panel (d) corresponds to a scan along the difference of the cavity detuning $\Delta$ for $\chi =3\pi/20$. The solid gray lines correspond to points at which both circulating intensities are equal, even though they are not simultaneous solutions to the homogeneous stationary state equations.
Particularly, the dotted red lines indicate the points at which the upper intensities of the two propagating fields match, corresponding to the curves at which asymmetric balance is fulfilled, as defined in Ref.~\citenum{GarbinPRR2020}. The dashed orange lines in (b-d) correspond to the points where both circulating intensities are equal and fulfill the stationary state equations. In panel (a), the solid black lines represent analytical results, while the remaining curves were obtained numerically. Our analytical tools are applicable for obtaining the solutions to the stationary states, but numerical tools were used to sort and compare them. We want to emphasize that these lines do not correspond to bifurcations. Rather, they exhibit additional structure within the solution space and serve to confirm and extend the findings of Ref.~\citenum{GarbinPRR2020}.

\section{Examples on resultant and Gröbner bases}

Here we present two examples to illustrate how the methods described in the main text, the resultant and Gröbner bases, can be used.

\subsection{Resultant}

Let us consider the following two (arbitrarily chosen) polynomials,
\begin{equation} \label{example1-eq}
    \begin{aligned}
        f(x,y) &=  x^3 - x^2 y^2, \\
        g(x,y) &= 2xy+y^2.
    \end{aligned}
\end{equation}
We want to find simultaneous solutions of $f(x,y)=0$ and $g(x,y)=0$. The resultant with respect to $x$, as defined in the Methods, is given by
\begin{equation}
    \textrm{res}(f,g,x) = \left| \begin{pmatrix}
 1 & 2 y & 0 & 0 \\
 -y^2 & y^2 & 2 y & 0 \\
 0 & 0 & y^2 & 2 y \\
 0 & 0 & 0 & y^2 \\
    \end{pmatrix}\right|  =  y^6 (1+ 2y),
\end{equation}
which is solved when $y = 0$, or $y=-1/2$.
By replacing these values of $y$ in Eq.~\eqref{example1-eq}, the corresponding values for~$x$, i.e. $x=0$ and $x = 1/4$, are found.
One alternative way would be to solve $f(x,y)=0$ for $y$, yielding $y = \pm \sqrt{x}$, and then inserting this in $f(x,y)=0$ yields \mbox{$\pm 2x^{3/2}+x = 0$, which is solved when $x = 0$ and $x = 1/4$}.
In more involved examples, this manual elimination method is either cumbersome, as one needs to keep track of all the different roots, or cannot be used when one considers polynomials of orders larger than four.

\subsection{Gröbner Basis}

For this example, let us consider the polynomials 
\begin{equation}\label{eq:example2}
    \begin{aligned}
        f(x,y,z) &= 2x+xz+ y+1,\\
        g(x,y,z) &= -x^2 + yz,\\
        h(x,y,z) &= -x-yz, 
    \end{aligned}
\end{equation}
and we try to find $f=0$, $g=0$ and $h=0$ simultaneously.

Like for resultants, many computer algebra systems implement algorithms determining Gröbner bases.
For our example, a possible Gröbner basis is
\begin{equation}\label{eq:Gröbner}
    G = \big(x^2+x,2 x+y-z+1,x z+z,x+z^2+z\big).
\end{equation}

The first element of this set is a univariate polynomial in~$x$, whose roots are given by $x = 0$ and \mbox{$x = -1$}.
Then for, e.g., \mbox{$x = -1$}, one repeats this procedure by determining a Gröbner basis for $f(-1,y,z)$, $g(-1,y,z)$ and $h(-1,y,z)$.
Alternatively, one notices that the last element of $G$ only depends on $z$ and the already known $x$, solve that for $z$, and then insert both results into the second element of $G$, solve, and ultimately find $(x,y,z) = (0,-1,0)$ and $(x,y,z)=(-1, (-1\pm\sqrt{5})/2, (-1+\sqrt{5})/2)$.

\end{document}